\documentclass[epj,twocolumn]{webofc}
\usepackage[varg]{txfonts}   
\newcommand{\rd}{\ensuremath{{\cal R}(D)}}
\newcommand{\rds}{\ensuremath{{\cal R}(D^*)}}
\newcommand{\rdrds}{\ensuremath{{\cal R}(D^{(*)})}}

\newcommand{\bbs}{\ensuremath{B_s\!-\!\overline{B}_s\,}}

\woctitle{Flavour changing and conserving processes}
\begin{document}
\title{New Physics in $B$ decays}
%
%

\author{Andreas Crivellin\inst{1} {\email{andreas.crivellin@cern.ch}}}

\institute{CERN Theory Division, CH--1211 Geneva 23, Switzerland }

\abstract{While the LHC did not observe direct evidence for physics beyond the standard model, indirect hints for new physics were uncovered in the flavour sector in the decays $B\to K^*\mu^+\mu^-$, $B\to K\mu^+\mu^-/B\to Ke^+e^-$, $B_s\to\phi\mu^+\mu^-$, $B\to D^{(*)}\tau\nu$ and $h\to\tau^\pm\mu^\mp$. Each observable deviates from the SM predictions at the $2-3\,\sigma$ level only, but combining all $b\to s\mu^+\mu^-$ data via a global fit, one finds $4-5\,\sigma$ difference for NP compared to the SM and combining $B\to D^{*}\tau\nu$ with $B\to D\tau\nu$ one obtains $3.9\,\sigma$.

While $B\to D^{(*)}\tau\nu$ and $h\to\tau\mu$ can be naturally explained by an extended Higgs sector, the $b\to s\mu^+\mu^-$ anomalies point at a $Z'$ gauge boson. However, it is also possible to explain $B\to D^{(*)}\tau\nu$ and $b\to s\mu^+\mu^-$ simultaneously with leptoquarks while their effect in $h\to\tau^\pm\mu^\mp$ is far too small to account for current data. Combining a 2HDM with a gauged $L_\mu-L_\tau$ symmetry allows for explaining the $b\to s\mu^+\mu^-$ anomalies in combination with $h\to\tau^\pm\mu^\mp$, predicting interesting correlations with $\tau\to3\mu$. In the light of these deviations from the SM we also discuss the possibilities of observing lepton flavour violating $B$ decays (e.g. $B\to K^{(*)}\tau^\pm\mu^\mp$ and $B_s\to\tau^\pm\mu^\mp$).}
\maketitle

\section{Introduction}

The LHC completed the standard model (SM) of particle physics by discovering the Higgs particle while no additional new particles have been observed so far. However, some indirect 'hints' for new physics (NP) in the flavor sector appeared in $B\to K^* \mu^+\mu^-$, $B_s\to\phi\mu^+\mu^-$, $R(K)=B\to K \mu^+\mu^-/B\to K e^+e^-$, $B\to D^{(*)}\tau\nu$ and $h\to\mu\tau$. 

Let us consider the current experimental and theoretical situation is some more detail. Concerning $b\to s\mu^+\mu^-$ transitions, already in 2013 LHCb reported deviations from the SM predictions~\cite{Egede:2008uy} (mainly in an angular observable called $P_5^\prime$~\cite{Descotes-Genon:2013vna}) in $B\to K^* \mu^+\mu^-$~\cite{Aaij:2013qta} with a significance of $2$--$3\,\sigma$ depending on the assumptions of hadronic uncertainties~\cite{Descotes-Genon:2014uoa,Altmannshofer:2014rta,Jager:2014rwa} being confirmed in 2015~\cite{LHCb:2015dla}. Also in the decay $B_s\to\phi\mu^+\mu^-$ \cite{Aaij:2015esa} LHCb uncovered differences compared to the SM prediction from lattice QCD \cite{Horgan:2013pva,Horgan:2015vla} of $3.5\,\sigma$ \cite{Altmannshofer:2014rta}. Furthermore, in 2014 LHCb~\cite{Aaij:2014ora} found indications for the violation of lepton flavour universality in
\begin{equation}
	R(K)=\frac{{\rm Br}[B\to K \mu^+\mu^-]}{{{\rm Br}[B\to K e^+e^-]}}=0.745^{+0.090}_{-0.074}\pm 0.036\,,
\end{equation}
in the range $1\,{\rm GeV^2}<q^2<6\,{\rm GeV^2}$ which disagrees with the theoretically clean SM prediction $R_K^{\rm SM}=1.0003 \pm 0.0001$~\cite{Bobeth:2007dw} by $2.6\,\sigma$. Combining these with other $b\to s$ observables, it is found that NP is preferred compared to the SM by $4-5\,\sigma$~\cite{Altmannshofer:2015sma,Descotes-Genon:2015uva}. Symmetry based solutions include NP in $C_9^{\mu\mu}$ (i.e. left-handed $b-s$ current and vectorial muon current) and $C_9^{\mu\mu}=-C_{10}^{\mu\mu}$ (i.e. left-handed $b-s$ and muon current).

Hints for lepton flavour universality violating NP in $B$ decays were observed for the first time by the BaBar collaboration in $B\to D^{(*)}\tau\nu$~\cite{Lees:2012xj} in 2012. Recently, these measurements have been confirmed by BELLE~\cite{Huschle:2015rga} and LHCb measured $B\to D^{*}\tau\nu$~\cite{Aaij:2015yra}. In summary, these experiments found for the ratios ${R}(D^{(*)})\equiv{\rm Br}(B\to D^{(*)} \tau \nu)/{\rm Br}(B\to D^{(*)} \ell \nu)$~\cite{Amhis:2014hma}:
\begin{align}
R(D)_{\rm EXP}\,&=\,0.391\pm0.041\pm0.028  \,,\\ 
R(D^*)_{\rm EXP}\,&=\,0.322\pm0.018\pm0.012  \,.
\end{align}
Comparing these measurements to the SM predictions~\cite{Fajfer:2012vx} $R_{\rm SM}(D)\,=\,0.297\pm0.017$, $R_{\rm SM}(D^*) \,=\,0.252\pm0.003$, we see that there is a combined discrepancy of $3.9\, \sigma$~\cite{Amhis:2014hma}. 

CMS recently also searched for the decay $h\to\tau\mu$ ~\cite{CMS:2014hha} finding a non-zero result of	${\rm Br} [h\to\mu\tau] = \left( 0.89_{-0.37}^{+0.40} \right)$ which disagrees by about $2.4 \,\sigma$ from 0, i.e. from the SM value. This is consistent with the less precise ATLAS measurement~\cite{Aad:2015gha} giving a combined significance of $2.6 \,\sigma$.

In these proceedings, we review NP models which can explain the deviations from the SM discussed above with focus on models with a $Z^\prime$ boson and/or additional Higgs doublets and also briefly mention models with leptoquarks.

\begin{figure*}[t]
\includegraphics[width=0.49\textwidth]{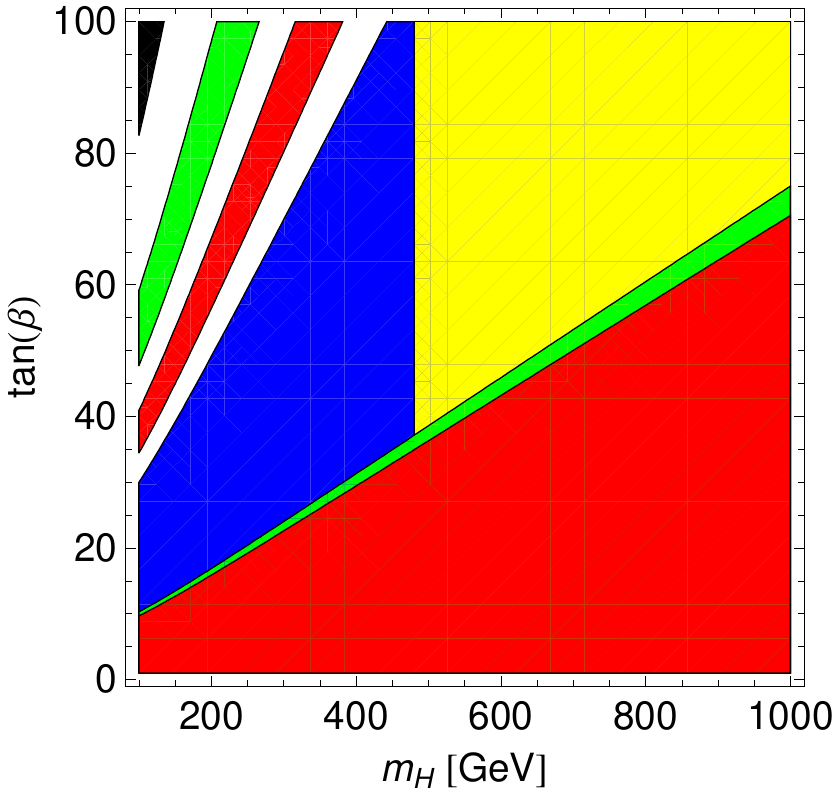}
\includegraphics[width=0.48\textwidth]{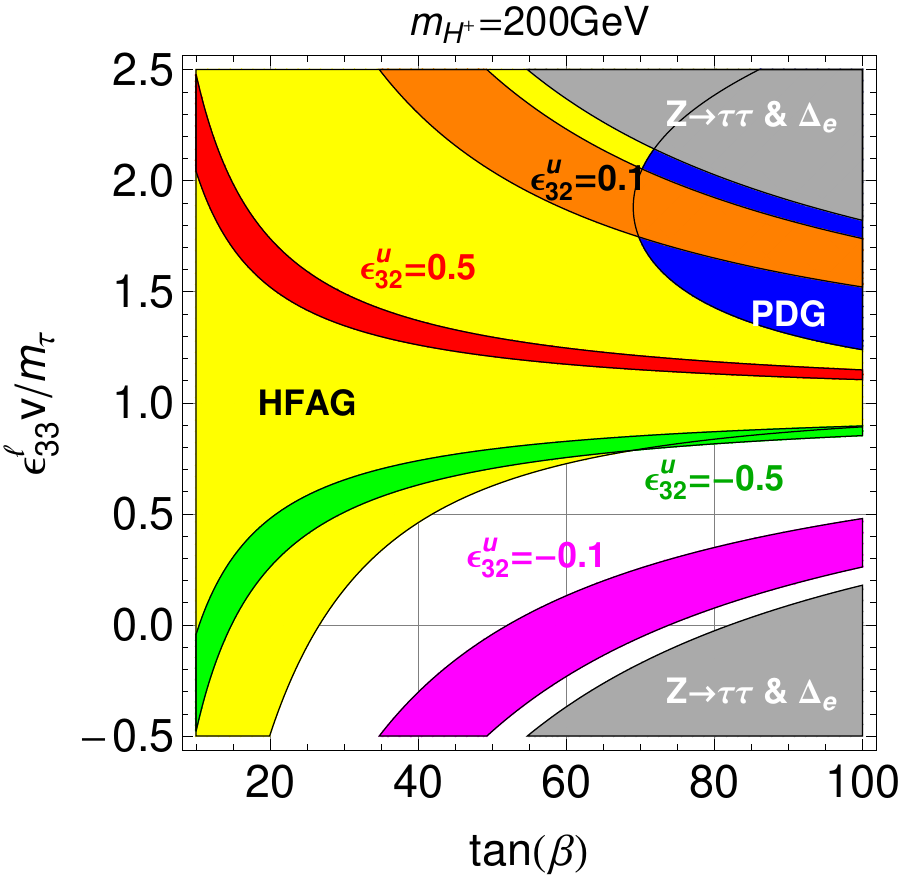}
\caption{
Left: Updated constraints on the 2HDM of type II parameter space. The regions compatible with experiment are shown (the regions are superimposed on each other): $b\to s\gamma$ (yellow) \cite{Misiak:2015xwa}, $B\to D\tau\nu$ (green), $B\to \tau \nu$ (red), $B_{s}\to \mu^{+} \mu^{-}$ (orange), $K\to \mu \nu/\pi\to \mu \nu$ (blue) and $B\to D^*\tau \nu$ (black). Note that no region in parameter space is compatible with all processes since explaining $B\to D^*\tau \nu$ would require very small Higgs masses and large values of $\tan\beta$ which is not compatible with the other observables. To obtain this plot, we added the theoretical uncertainty of the SM linearly on the top of the $2 \, \sigma$ experimental error. 
Right: Allowed regions in the $\tan\beta$--$v/m_\tau\epsilon^\ell_{33}$ plane from \rdrds{} and $\tau\to\mu\nu\nu$ at the $2\,\sigma$ level in the perturebed 2HDM of type X~\cite{Crivellin:2015hha}. The yellow region is allowed by $\tau\to\mu\nu\nu$ using the HFAG result for $m_H=30\,$GeV and $m_A=200\,$GeV, while the (darker) blue one is the allowed region using the PDG result. The red, orange, green, and magenta bands correspond to the allowed regions by \rdrds{} for different values of $\epsilon^u_{32}$. The gray region is excluded by $Z\to \tau\tau$ and $\tau\to e \nu\nu$. For $m_H\simeq m_A$ the allowed regions from $\tau\to\mu\nu\nu$ would be slightly larger.
\label{fig:2HDMII}}
\end{figure*}

\section{Tauonic $B$ Decays}

Due to the heavy tau lepton in the final state, these decays are sensitive to charged Higgses~\cite{Krawczyk:1987zj}. A 2HDM of type II (like the MSSM at tree-level) cannot explain the deviations from the SM in tauonic $B$ decays (due to the necessarily destructive interference) without violating bounds from other observables~\cite{Crivellin:2013wna} (see left plot in Fig.~\ref{fig:2HDMII}). However, a 2HDM with generic Yukawa coupling (i.e. type III) can account for $B\to D\tau\nu$ and $B\to D^*\tau\nu$ simultaneously, respecting the constraints from all other observables, if the coupling of a right-handed charm to the third generation quark doublet ($\epsilon^u_{32}$) is large~ \cite{Crivellin:2012ye,Crivellin:2015hha}. 

\begin{figure*}[t]
\centering
\includegraphics[width=0.45\textwidth]{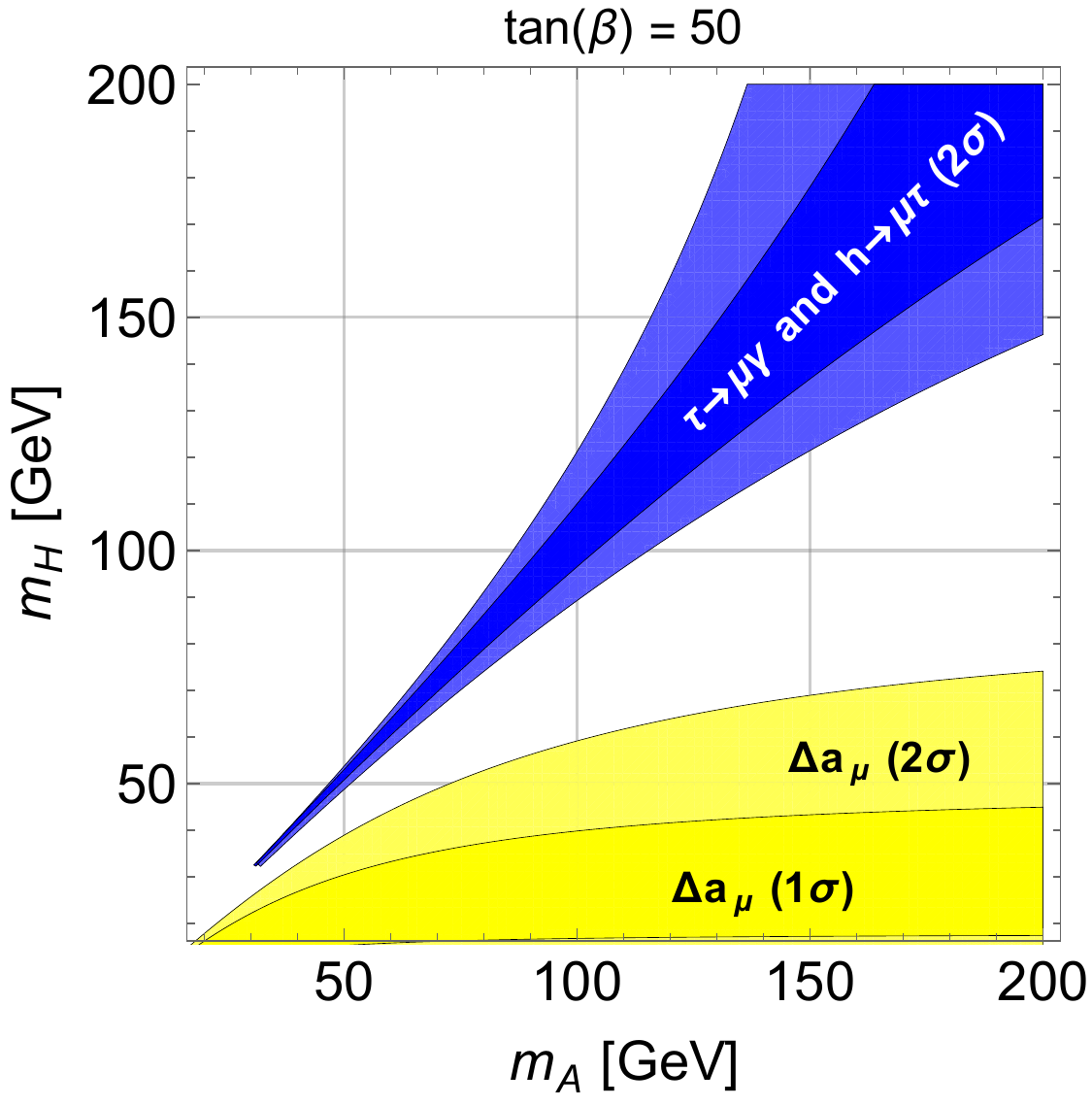}
\includegraphics[width=0.45\textwidth]{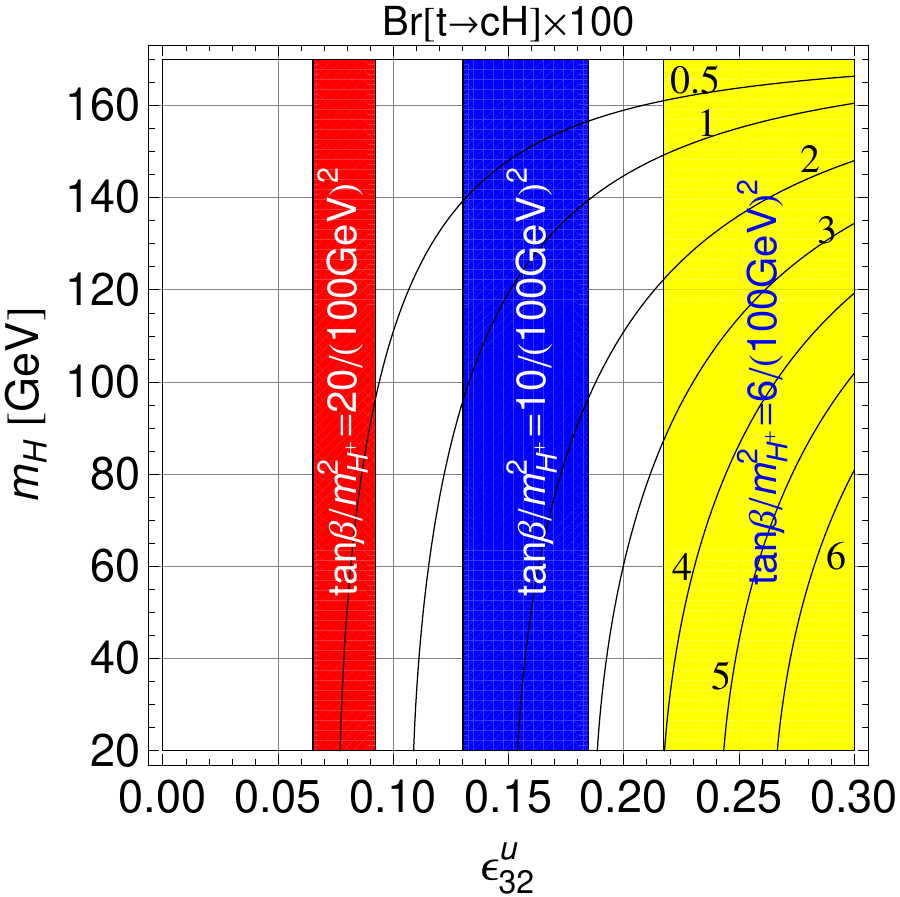}
\caption{Left: Red, green, and yellow are the allowed regions in the $m_A$--$m_H$ plane from $(g-2)_\mu$ at the $2\sigma$ level for $\tan\beta=50$, $m_{H^+}=200\,$GeV, $\cos(\alpha-\beta)=0.1$ and different values of $\epsilon^\ell_{33}$ in the perturbed 2HDM X. Blue is the allowed region (again at $2\sigma$) from $\tau\to\mu\gamma$ and $h\to\mu\tau$ for $\epsilon^\ell_{33} = 2m_\tau/v$ and $\cos(\alpha-\beta)=0.1$, light blue corresponds to $\cos(\alpha-\beta)=0.2$. The allowed region for $\Delta a_\mu$ covers the three possibilities $\epsilon^\ell_{32}\neq0$, $\epsilon^\ell_{32}=\epsilon^\ell_{23}\neq0$ and $\epsilon^\ell_{32}=-\epsilon^\ell_{23}\neq0$, since the latter ones can give $m_\tau/m_\mu$ enhanced one-loop contributions. However, the effects turn out to be small, as $\epsilon^\ell_{32,32}$ is stringently constrained from $\tau\to\mu\gamma$. In addition, we checked that the effect of $\lambda_H$ is very small. The white regions are not compatible with experiment at the $2\,\sigma$ level.
Right: The contour lines denote ${\rm BR}(t\to H c)\times 100$ as a function of $\epsilon^u_{32}$ and $m_H$. The colored regions are allowed by \rd{} and \rds{} for different values of $\tan\beta/m_{H^+}^2$. Note that $H$ is required to be quite light if one aims at explaining the anomalous magnetic moment of the muon.
\label{topHc}}
\end{figure*}

Here, two variants are phenomenologically possible: in the limit of vanishing non-standard couplings, the type III model could reduce either to type II (like the MSSM at tree-level) or type X (leptospecific). While the low energy constraints on the type II model are quite stringent (see left plot in Fig.~\ref{fig:2HDMII}) and it is also challenged by $A\to\tau\tau$ searches~\cite{CMS:2013hja}, the type X model is only weakly constrained (see for example Ref.~\cite{Branco:2011iw} for a review). Therefore, we will focus on the type X model as a solution~\cite{Crivellin:2015hha} which has also the advantage of providing a possible explanation for the anomalous magnetic moment of the muon~\cite{Broggio:2014mna,Wang:2014sda,Abe:2015oca} and $\tau\to\mu\nu\nu$. In the right plot in Fig.~\ref{fig:2HDMII} we show which regions in parameter space can account for the experimental data. As usual, $\tan\beta$ is the ratio of the two vacuum expectation values and $m_H$ ($m_A$) refer to the additional neutral CP-even (CP-odd) Higgs mass. Interestingly, requiring an explanation for the anomalous magnetic moment of the muon without violating bounds from $\tau\to\mu\nu\nu$ enforces $m_H\ll m_A$ (see left plot of Fig.~\ref{topHc}). Together with the large top-charm coupling induced by $\epsilon^u_{32}$ (necessary for $R(D^{(*)})$) sizable branching ratios for $t\to Hc$ (see right plot of Fig.~\ref{topHc}) are predicted which are well within the reach of the LHC.

Alternative explanations involve leptoquarks~\cite{Fajfer:2012jt,Sakaki:2013bfa,Alonso:2015sja,Freytsis:2015qca,Calibbi:2015kma,Bauer:2015knc} or R-parity violating SUSY~\cite{Deshpande:2012rr}.

\section{Anomalies in $b\to s\mu^+\mu^-$}

A rather large contribution to operator $(\overline{s}\gamma_\alpha P_L b)(\overline{\mu}\gamma^\alpha \mu)$, as preferred by the model independent fit~\cite{Altmannshofer:2015sma,Descotes-Genon:2015uva}, can be achieved in models containing a heavy $Z^\prime$ gauge boson (see Refs.~\cite{Gauld:2013qba,Altmannshofer:2014cfa} for early attempts to explain this anomaly). If one aims at explaining $R(K)$ as well, a contribution to $C_9^{\mu\mu}$ involving muons, but not to $C_9^{ee}$ with electrons is necessary~\cite{Alonso:2014csa,Hiller:2014yaa,Ghosh:2014awa}. This is naturally the case in models with gauged muon minus tauon number ($L_\mu-L_\tau$)~\cite{Altmannshofer:2014cfa,Crivellin:2015mga,Crivellin:2015lwa}\footnote{$Z^\prime$ bosons with the desired couplings can also be obtained in other $Z^\prime$ models \cite{Niehoff:2015bfa,Sierra:2015fma,Celis:2015ara,Belanger:2015nma,Falkowski:2015zwa,Carmona:2015ena}. For an extensive analysis of $Z^\prime$ models prior to the apparence of the anomalies see for example~\cite{Buras:2012jb,Buras:2013dea,Buras:2013qja}.}. Alternative explanations are again models with leptoquarks~\cite{Gripaios:2014tna,Becirevic:2015asa,Varzielas:2015iva,Alonso:2015sja,Calibbi:2015kma,Bauer:2015knc}. 

In $Z^\prime$ models the couplings to quarks can be written generically as
\begin{align}
L \cup g'\left({{{\bar d}}_i}{\gamma ^\mu }{P_L}{{d}_j}{Z'_\mu }\Gamma_{ij}^{{d}L} + {{{\bar d}}}_i{\gamma ^\mu }{P_R}{{d}_j}{Z'_\mu }\Gamma_{ij}^{{d}R}\right)\,.
\end{align}
where $g^\prime$ is the new $U^\prime(1)$ gauge coupling constant. Unavoidable tree-level contributions to \bbs are generated which constrain the coupling to muons to be much larger than the one to $\bar s b$. In the left plot in Fig.~\ref{fig:HiggsPlot} the regions in the $\Gamma^L_{sb}$--$\Gamma^R_{sb}$ plane are shown which are in agreement with \bbs mixing and $b\to s\mu^+\mu^-$ data within $2\,\sigma$. Note that in the symmetry limit $\Gamma^R_{sb}=0$, \bbs mixing puts a upper bound on $\Gamma^L_{sb}$.

\subsection{$h\to \tau\mu$}

Lepton flavour violating couplings of the SM Higgs are induced by a single operator up to dim-6. Considering only this operator ${\rm Br}[h\to\mu\tau]$ can be up to $10\%$~\cite{Davidson:2012ds,Kopp:2014rva}. However, it is in general difficult to get dominant contributions to this operator in a UV complete model, as for example in models with vector-like leptons~\cite{Falkowski:2013jya} or leptoquarks~\cite{Dorsner:2015mja,Altmannshofer:2015esa}. Therefore, among the several attempts to explain this $h\to\mu\tau$ observation, most of them are relying on models with extended Higgs sectors~\cite{Sierra:2014nqa,Dorsner:2015mja}. One particularly elegant solution employs a two-Higgs-doublet model (2HDM) with gauged $L_\mu-L_\tau$~\cite{Heeck:2014qea}.

\section{Simultaneous explanation of $b\to s\mu\mu$ and $h\to \tau\mu$ and predictions for $\tau\to3\mu$}

In \cite{Crivellin:2015mga,Crivellin:2015lwa} two models with gauged $L_\mu-L_\tau$ symmetry were presented which can explain $h\to \tau\mu$ simultaneously with the anomalies in $b\to s\mu\mu$ data (including $R(K)$) giving rise to interesting correlated effects in $\tau\to3\mu$. While in both models the $Z'$ couplings to leptons originate from a gauged $L_\mu-L_\tau$ symmetry, the coupling to quarks is either generated effectively via heavy vector-like quarks charged under $L_\mu-L_\tau$ or directly by assigning horizontal changes to baryons\footnote{For pioneering work on horizontal $U(1)$ gauge symmetries see Ref.~\cite{Terazawa:1976xx}.}.

\subsection{2 Higgs doublets with vector-like quarks}

\begin{figure*}
\includegraphics[width=0.463\textwidth]{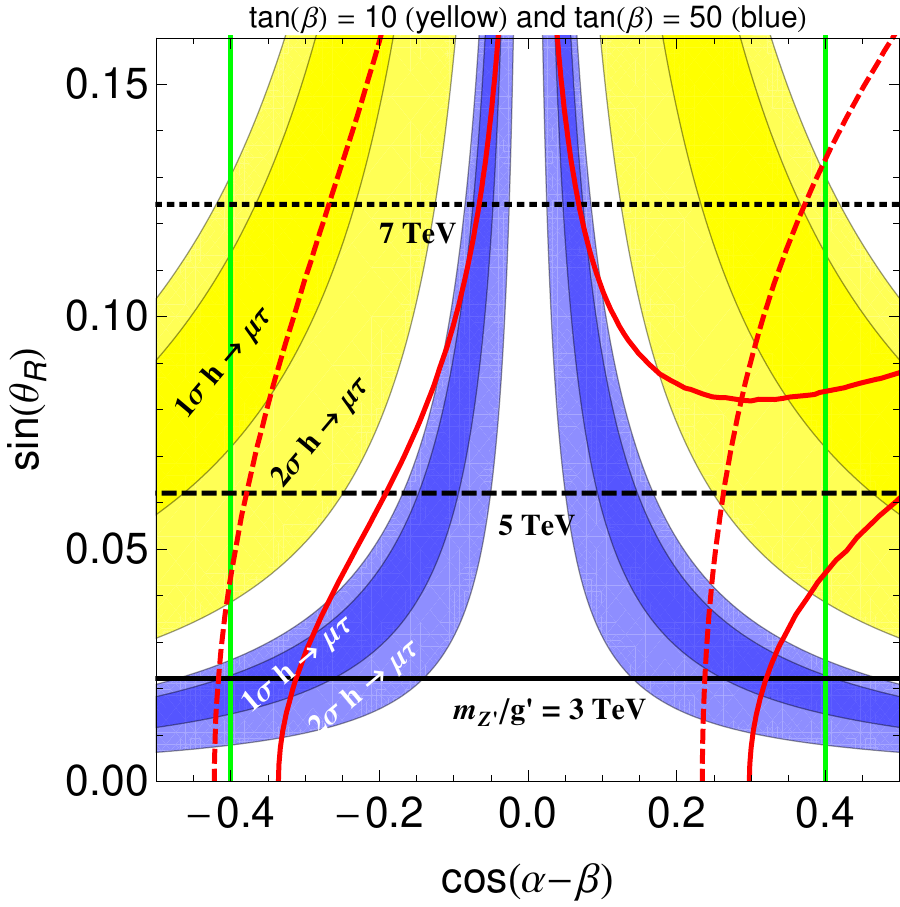} \hspace{3ex}
\includegraphics[width=0.45\textwidth]{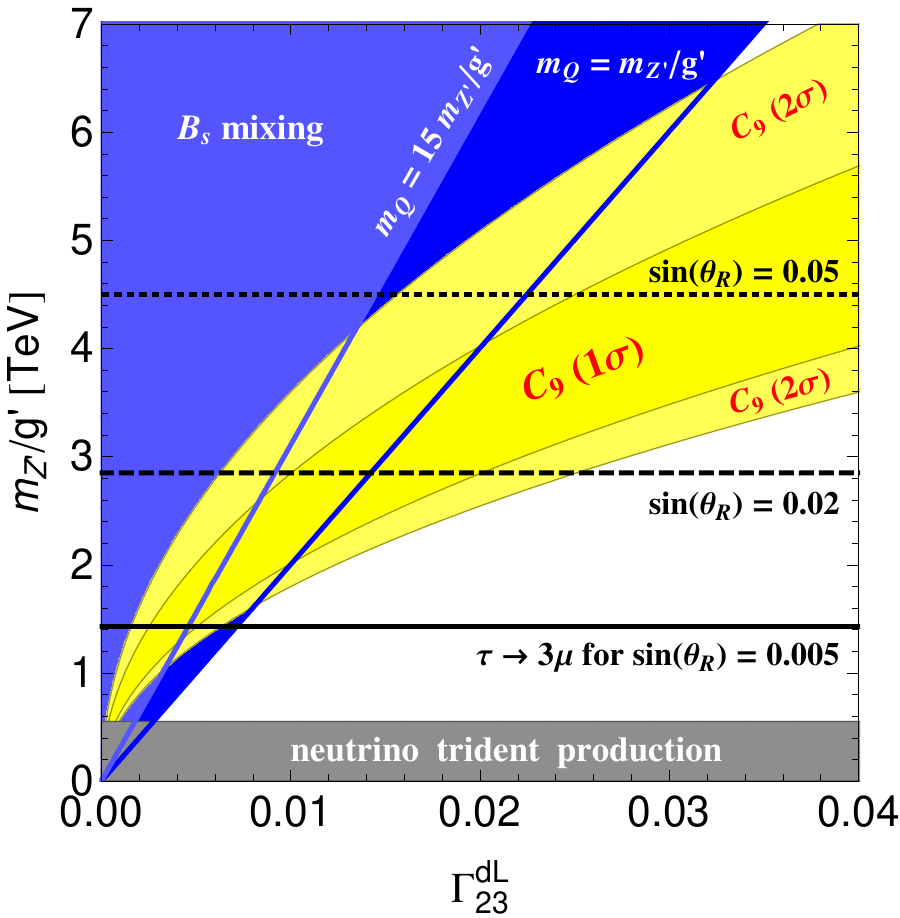}
\caption{ 
Left: Allowed regions in the $\Gamma^{L}_{sb}/M_{Z^\prime}-\Gamma^{R}_{sb}/M_{Z^\prime}$ plane for $g^\prime=1$ from $B_s$-$\overline{B}_s$ mixing (blue), and from the $C^{\mu\mu}_9-C^{(\prime)\mu\mu}_9$ fit of Ref.~\cite{Altmannshofer:2014rta} to $b\to s\mu^+\mu^-$ data, with $\Gamma^V_{\mu\mu}=\pm 1$ (red), $\Gamma_{\mu\mu}^V=\pm 0.5$ (orange) and  $\Gamma^V_{\mu\mu}=\pm 0.3$ (yellow). Note that the allowed regions with positive (negative) $ \Gamma^{L}_{sb}$ correspond to positive (negative) $\Gamma^V_{\mu\mu}$.
Right: Allowed regions in the $\Gamma^{dL}_{23}$--$m_{Z^\prime}/g^\prime$ plane from $b\to s\mu^+\mu^-$ data (yellow) and $B_s$ mixing (blue). For $B_s$ mixing (light) blue 
corresponds to ($m_Q=15 m_{Z^\prime}/g^\prime$) $m_Q=m_{Z^\prime}/g^\prime$. The horizontal lines denote the lower bounds on $m_{Z^\prime}/g^\prime$ from $\tau\to3\mu$ for $\sin(\theta_R)=0.05,\; 0.02,\; 0.005$. The gray region is excluded by NTP.\label{fig:HiggsPlot}}
\end{figure*}

In a 2HDM with a gauged $U(1)_{L_\mu-L_\tau}$ symmetry~\cite{Heeck:2014qea}, $L_\mu-L_\tau$ is broken spontaneously by the vacuum expectation value of a scalar $\Phi$ (being singlet under the SM gauge group) with $Q^{\Phi}_{L_\mu-L_\tau}=1$, leading to the $Z'$ mass $m_{Z'} = \sqrt2 g' \langle\Phi\rangle \equiv g' v_\Phi$. Two Higgs doublets are introduced which break the electroweak symmetry: $\Psi_1$ with $Q^{\Psi_1}_{L_\mu-L_\tau}=-2$ and $\Psi_2$ with $Q^{\Psi_2}_{L_\mu-L_\tau}=0$. Therefore, $\Psi_2$ gives masses to quarks and leptons while $\Psi_1$ couples only off-diagonally to $\tau\mu$:
\begin{align}
\mathcal{L}_Y \ \supset\ &-\overline{\ell}_f Y^\ell_{i}\delta_{fi} \Psi_2 e_i - \xi_{\tau\mu} \overline{\ell}_3 \Psi_1 e_2  \nonumber \\&-\overline{Q}_f Y^u_{fi} \tilde{\Psi}_2 u_i - \overline{Q}_f Y^d_{fi} \Psi_2 d_i + \mathrm{h.c.}\,.
\label{eq:yukawas}
\end{align}
Here $Q$ ($\ell$) is the left-handed quark (lepton) doublet, $u$ ($e$) is the right-handed up-quark (charged-lepton) and $d$ the right-handed down quark while $i$ and $f$ label the three generations and the tilde signals charge conjugation. The scalar potential is the one of a $U(1)$-invariant 2HDM~\cite{Branco:2011iw} with additional couplings to the SM-singlet $\Phi$. We defined as usual $\tan\beta = \langle \Psi_2\rangle/\langle \Psi_1\rangle$ and $\alpha$ is the mixing angle between the neutral CP-even components of $\Psi_1$ and $\Psi_2$ (see for example~\cite{Branco:2011iw}). Therefore, quarks and gauge bosons have standard type-I 2HDM couplings to the scalars. The only deviations from the type I model are in the lepton sector: while the Yukawa couplings $Y^\ell_{i}\delta_{fi}$ of $\Psi_2$ are forced to be diagonal by the ${L_\mu-L_\tau}$ symmetry, $\xi_{\tau\mu}$ gives rise to an off-diagonal entry in the lepton mass matrix:
\begin{equation}
m^\ell_{fi}= \frac{v}{\sqrt{2}}\begin{pmatrix}
y_e\sin\beta &0&0\\
0& y_\mu \sin\beta& 0\\
0&\xi_{\tau\mu} \cos\beta& y_\tau \sin\beta
\end{pmatrix} .
\end{equation}
It is this $\tau$--$\mu$ element that leads to the LFV couplings of $h$ and $Z'$. The mass basis for the charged leptons is obtained by rotating $(\mu_R,\tau_R)$ and $(\mu_L,\tau_L)$ with the angles $\theta_R$ and $\theta_L$. A non-vanishing angle $\theta_R$ not only gives rise to the LFV decay $h\to\mu\tau$ due to the coupling
\begin{equation}
\frac{m_\tau}{v}\frac{\cos(\alpha-\beta)}{\cos(\beta)\sin(\beta)}\sin(\theta_R)\cos(\theta_R) \bar\tau P_R\mu h\equiv \Gamma^{h}_{\tau\mu}\bar\tau P_R\mu h\,,\label{h0taumu}
\end{equation}
in the Lagrangian, but also leads to off-diagonal $Z'$ couplings to right-handed leptons
\begin{align}
g^\prime Z^\prime_\nu \, (\overline{\mu}, \overline{\tau})
\begin{pmatrix}
 \cos 2\theta_R& \sin 2\theta_R\\
\sin 2\theta_R& - \cos 2\theta_R
\end{pmatrix} \gamma^\nu P_R 
\begin{pmatrix}
 \mu\\
\tau
\end{pmatrix} ,
\end{align}
while the left-handed couplings are to a good approximation flavour conserving. In order to explain the observed anomalies in the $B$ meson decays, a coupling of the $Z'$ to quarks is required as well, not inherently part of $L_\mu-L_\tau$ models (aside from the kinetic $Z$--$Z'$ mixing, which is assumed to be small). Following Ref.~\cite{Altmannshofer:2014cfa}, effective couplings of quarks to the $Z^\prime$ are generated by heavy vector-like quarks \cite{Langacker:2008yv} charged under $L_\mu-L_\tau$. As a result, the couplings of the $Z^\prime$ to quarks are in principle free parameters. In the limit of decoupled vector-like quarks with the quantum numbers of right-handed quarks, only $C_9$ is generated, giving a very good fit to data. The results are shown in the right plot of Fig.~\ref{fig:HiggsPlot} depicting that for small values of $\Gamma^L_{sb}$ and $\theta_R$, $b\to s\mu^+\mu^-$ data can be explained without violating bounds from $B_s-\overline{B}_s$ mixing or $\tau\to3\mu$. In the left plot of Fig.~\ref{fig:vevplot} the correlations of $b\to s\mu^+\mu^-$ and $h\to\tau\mu$ with $\tau\to3\mu$ are shown, depicting that consistency with $\tau\to3\mu$ requires large values of $\tan\beta$ (not being in conflict with any data as the decoupling limit (i.e. $\xi_{\tau\mu}=0$) is a type I model) and future searches for $\tau\to3\mu$ are promising to yield positive results.

\subsection{Horizontal charges for quarks}

\begin{figure*}[t]
\includegraphics[width=0.41\textwidth]{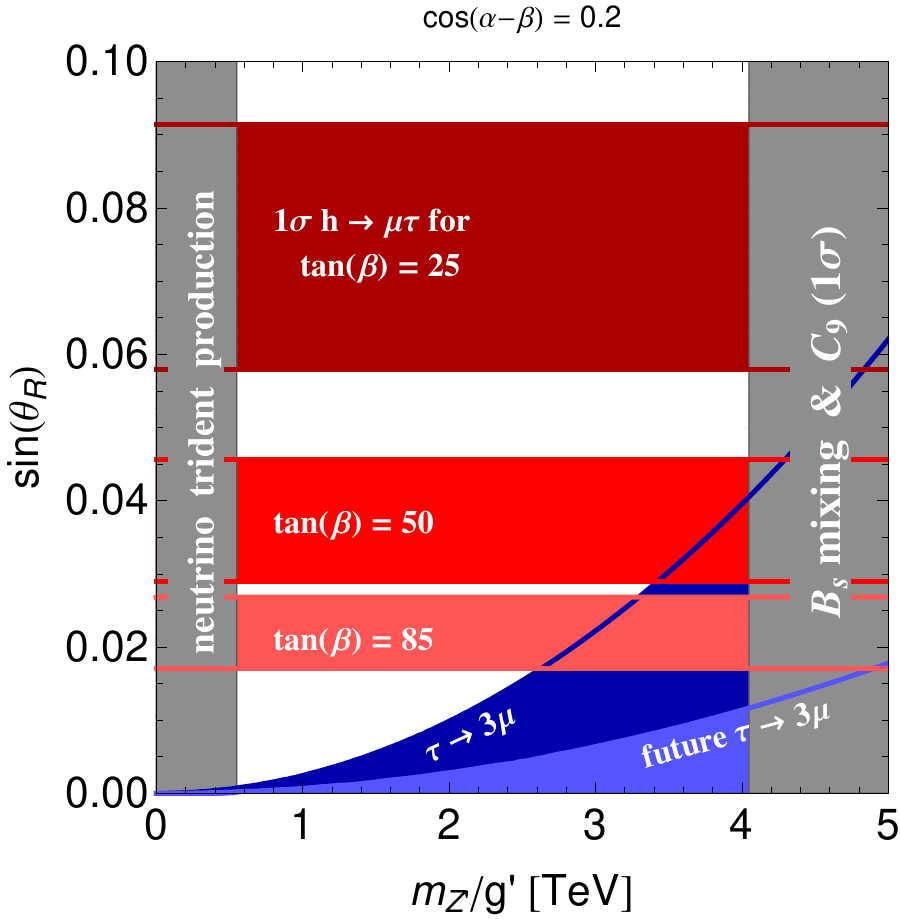}
\includegraphics[width=0.59\textwidth]{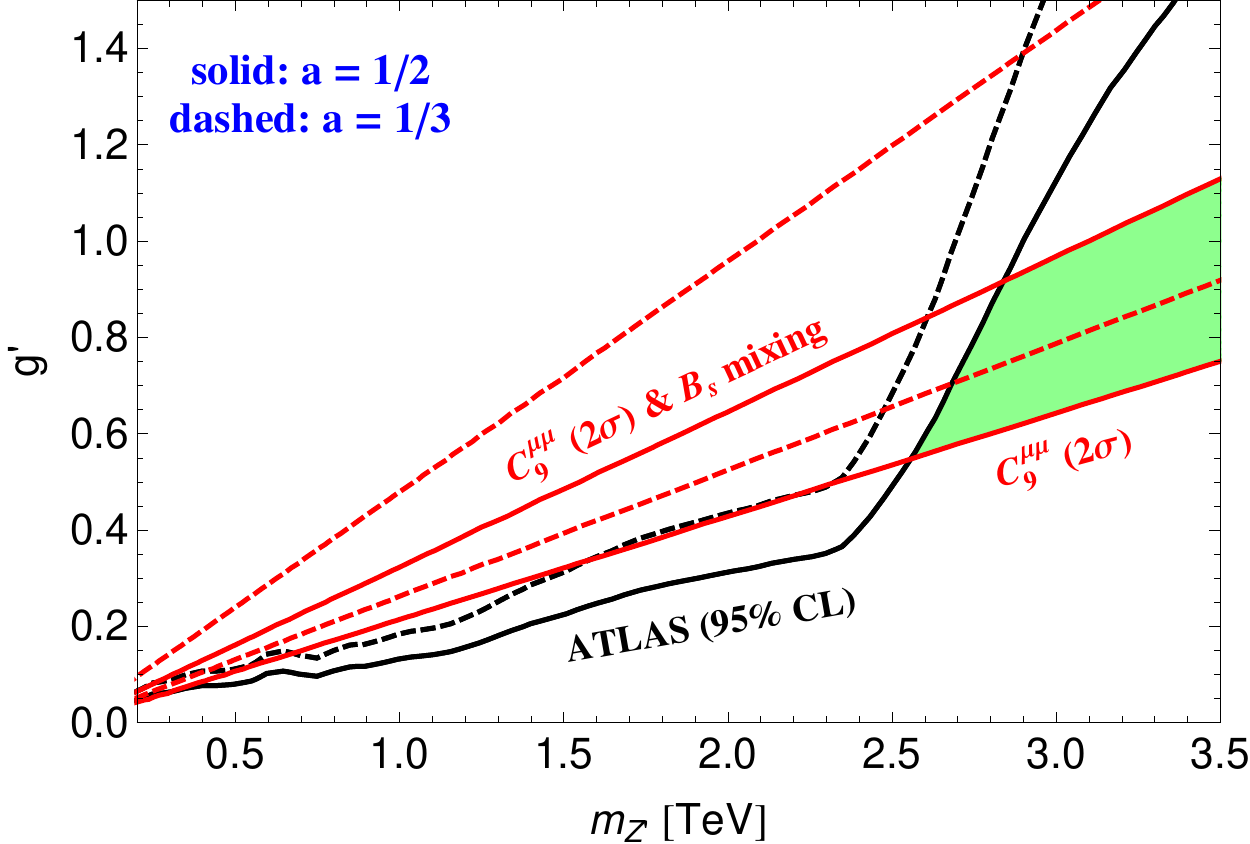}
\caption{Left: Allowed regions in the $m_{Z'}/g'$--$\sin (\theta_R)$ plane: the horizontal stripes correspond to $h\to\mu\tau$ ($1\sigma$) for $\tan\beta=85,\,50,\,25$ and $\cos (\alpha-\beta)=0.2$, (light) blue stands for (future) $\tau\to 3\mu$ limits at $90\%$~C.L. The gray regions are excluded by NTP or $B_s$--$\overline{B}_s$ mixing in combination with the $1\,\sigma$ range for $C_9$. \newline
Right: Limits on $q\overline{q}\to Z' \to \mu\overline{\mu}$ from ATLAS~\cite{Aad:2014cka} (black, allowed region down right) and the $2\sigma$ limits on $C_9^{\mu\mu}$ to accommodate $b\to s\mu^+\mu^-$ data (red, allowed regions inside the cone). Solid (dashed) lines are for $a=1/2$ ($a=1/3$). For $a =1/2$, the green shaded region is allowed (similar for $a= 1/3$ using the dashed bounds).}
\label{fig:vevplot}
\end{figure*}

In order to avoid the introduction of vector-like quarks, one can introduce flavour-dependent charges to quarks as well \cite{Crivellin:2015lwa}. Here, the first two generations should have the same charges in order to avoid very large effects in $K$--$\overline{K}$ or $D$--$\overline{D}$ mixing, generated otherwise unavoidably due to the breaking of the symmetry necessary to generate the measured Cabibbo angle of the CKM matrix. If we require in addition the absence of anomalies, we arrive at the following charge assignment for baryons	$Q'(B)= (-a,\,-a,\,2a )$.
Here $a \in {\cal Q}$ is a free model parameter with important phenomenological implications. 
In this model, at least one additional Higgs doublet which breaks the flavour symmetry in the quark sector is required, and one more is needed if one attempts to explain $h\to\tau\mu$. In case the mixing among the doublets is small, the correlations among $h\to\tau\mu$, $b\to s\mu^+\mu^-$ and $\tau\to 3\mu$ are the same is in the model with vector-like quarks discussed in the last subsection and shown in the left plot of Fig.~\ref{fig:vevplot}.

The low-energy phenomenology is rather similar to the one of the model with vector like quarks considered in the last section, but the contributions to $B_s-\overline{B}_s$ mixing are directly correlated to $B_d-\overline{B}_d$ and $K-\overline{K}$ mixing as all flavour violation is due to CKM factor. However, concerning direct LHC searches, the implications are very different, as the $Z^\prime$ boson can be directly produced on-shell as a resonance in $p\bar p$ collisions since it couples to quarks of the first generation. The resulting strong bounds are shown in right plot of Fig.~\ref{fig:vevplot} where they are compared to the allowed regions from $B_s-\overline{B}_s$ mixing and $b\to s\mu^+\mu^-$ data for different values of $a$.

\section{Simultaneous explanation of $b\to s\mu^+\mu^-$ data and $R(D^{(*)})$}

\subsection{Effective Operators}

A scenario with left-handed currents only gives a good fit to $b\to s\mu^+\mu^-$ data~\cite{Altmannshofer:2015sma,Descotes-Genon:2015uva}. In such a scenario $SU(2)_L$ relations are necessarily present and connect charged to neutral currents. These relations are automatically taken into account once gauge invariant operators are considered. There are only two left-handed 4-fermion operators in the effective Lagrangian
\begin{equation}
	{\cal L_{\rm dim6}}=\dfrac{1}{\Lambda^2}\sum{O_X C_X}\,,
\end{equation}
where $\Lambda$ is the scale of NP, which can contribute to $b\to s\ell\ell$ transitions at tree-level~\cite{Buchmuller:1985jz,Grzadkowski:2010es}:
\begin{align}
\label{Q1-Q3}
Q_{\ell q}^{\left( 1 \right)} = \left( {\bar L{\gamma ^\mu }L} \right)\left( {\bar Q{\gamma _\mu }Q} \right)\,,~~
Q_{\ell q}^{\left( 3 \right)} = \left( {\bar L{\gamma ^\mu }{\tau _I}L} \right)\left( {\bar Q{\gamma _\mu }{\tau ^I}Q} \right)\,.
\end{align}
Here $L$ is the lepton doublet, $Q$ the quark doublet and the flavour indices are not explicitly shown. Writing these operators in terms of their $SU(2)_L$ components (i.e.~up-quarks, down-quarks, charged leptons and neutrinos) we find for the terms relevant for the processes discussed in the last section (before EW symmetry breaking)
\begin{align}
{\cal L} \supset & \frac{{C_{ijkl}^{\left( 1 \right)}}}{{{\Lambda ^2}}}\left( {{{\bar \ell }_i}{\gamma ^\mu }{P_L}{\ell _j}{{\bar d}_k}{\gamma _\mu }{P_L}{d_l} + {{\bar \nu }_i}{\gamma ^\mu }{P_L}{\nu _j}{{\bar d}_k}{\gamma _\mu }{P_L}{d_l}} \right) \nonumber\\+
 &\frac{C_{ijkl}^{\left( 3 \right)}}{{{\Lambda ^2}}}\left( {2{{\bar \ell }_i}{\gamma ^\mu }{P_L}{\nu _j}{{\bar u}_k}{\gamma _\mu }{P_L}{d_l} - {{\bar \nu }_i}{\gamma ^\mu }{P_L}{\nu _j}{{\bar d}_k}{\gamma _\mu }{P_L}{d_l}}\right.\nonumber\\ 
 &+\left.{ {{\bar \ell }_i}{\gamma ^\mu }{P_L}{\ell _j}{{\bar d}_k}{\gamma _\mu }{P_L}{d_l}} \right)\,,
\end{align}
where $C^{(1,3)}_{ijkl}$ are the dimensionless coefficients of the operators of Eq.~(\ref{Q1-Q3}). After EW symmetry breaking the following redefinitions of the fields are performed in order to render the mass matrices diagonal
\begin{equation}
{d_L} \to D_{}^\dag {d_L},\;{u_L} \to U_{}^\dag {u_L},\;\;{\ell _L} \to L_{}^\dag {\ell _L},\;\nu  \to L_{}^\dag \nu \,.
\end{equation}
Defining
\begin{equation}
\lambda^{(1,3)}\tilde X_{ij}^{\left( {1,3} \right)}\tilde Y_{kl}^{\left( {1,3} \right)} = L_{i'i}^*{L_{j'j}}D_{k'k}^*{D_{l'l}}C_{i'j'k'l'}^{\left( {1,3} \right)}\,,
\end{equation}
where $\lambda^{(1,3)}$ are overall constants, we finally obtain
\begin{align}
C_9^{ij} =&  -C_{10}^{ij}
\nonumber 
\\ =& \frac{\pi }{{\sqrt 2 {\Lambda ^2}{G_F}\alpha {V_{tb}}V_{ts}^*}}\left( {{\lambda ^{(1)}}\tilde X_{ij}^{(1)}\tilde Y_{23}^{(1)} + {\lambda ^{(3)}}\tilde X_{ij}^{(3)}\tilde Y_{23}^{(3)}} \right)\nonumber\\
C_L^{ij} 
=& 
\frac{\pi }{{\sqrt 2 {\Lambda ^2}{G_F}\alpha {V_{tb}}V_{ts}^*}}\left( {{\lambda ^{(1)}}\tilde X_{ij}^{(1)}\tilde Y_{23}^{(1)} - {\lambda ^{(3)}}\tilde X_{ij}^{(3)}\tilde Y_{23}^{(3)}} \right)\nonumber\,,
\end{align}
\begin{align}
C_{L\;ij}^{cb} &=  - {\frac{\lambda^{(3)}}{{\sqrt 2 {\Lambda ^2}{G_F}}}}\frac{{{{\tilde X}_{ij}^{(3)}}}}{{{V_{cb}}}}\sum\limits_k {\left( {{V_{2k}}{{\tilde Y}_{k3}^{(3)}}} \right)}\,,\label{C9C10CL}
\end{align}
for the Wilson coefficients relevant for $b\to s\mu^+\mu^-$,  $B\to K^{(*)}\nu\bar{\nu}$ and $B\to D^{(*)}\tau\nu$ respectively. Note that in the limit $C^{(1)}=C^{(3)}$ the contribution to $B\to K^{(*)}\nu\bar{\nu}$ vanishes. 

\begin{figure}[t]
\begin{center}
\begin{tabular}{cp{7mm}c}
\includegraphics[width=0.43\textwidth]{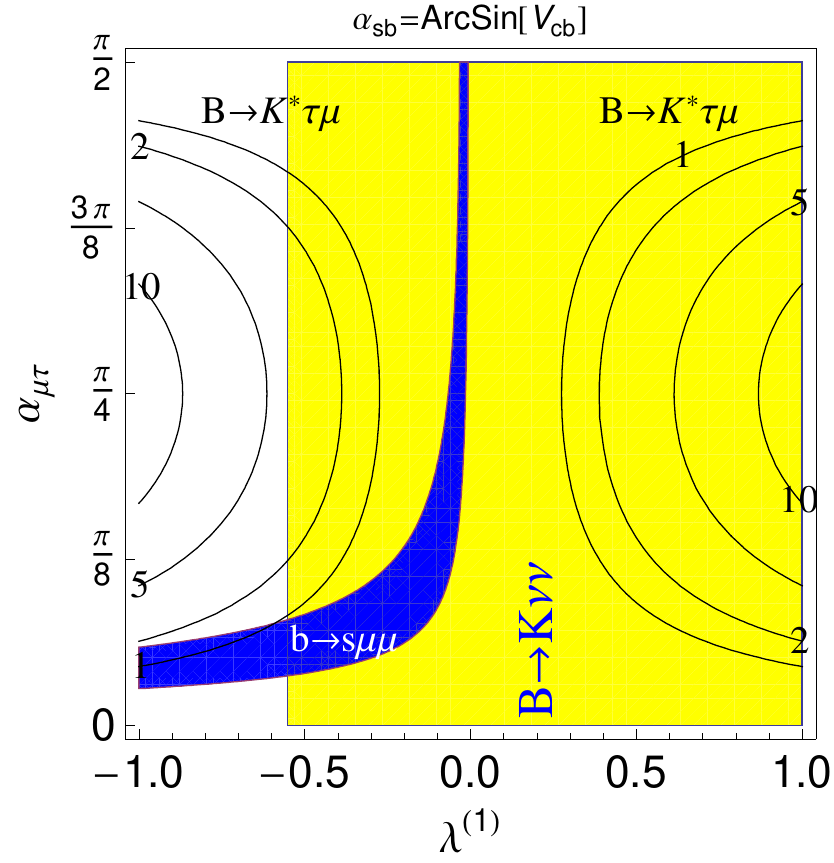} &&
\end{tabular}
\end{center}
\caption{Allowed regions in the $\lambda^{(1)}$--$\alpha_{\mu\tau}$ plane from
 $b\to s \mu^+\mu^-$ data (blue) and $B\to K \nu\bar{\nu}$ (yellow) for $\alpha_{sb}={\rm ArcSin}[V_{cb}]$ and $\Lambda=1\,$TeV. Note that here changing $\alpha_{sb}$ only has the effect of an overall scaling of $\lambda^{(1)}$. The contour lines denote ${\rm Br}[B\to K^*\tau\mu]$ in units of $10^{-6}$.
\label{Plot1}}
\end{figure}

\begin{figure*}[t]
\begin{center}
\begin{tabular}{cp{7mm}c}
\includegraphics[width=0.32\textwidth]{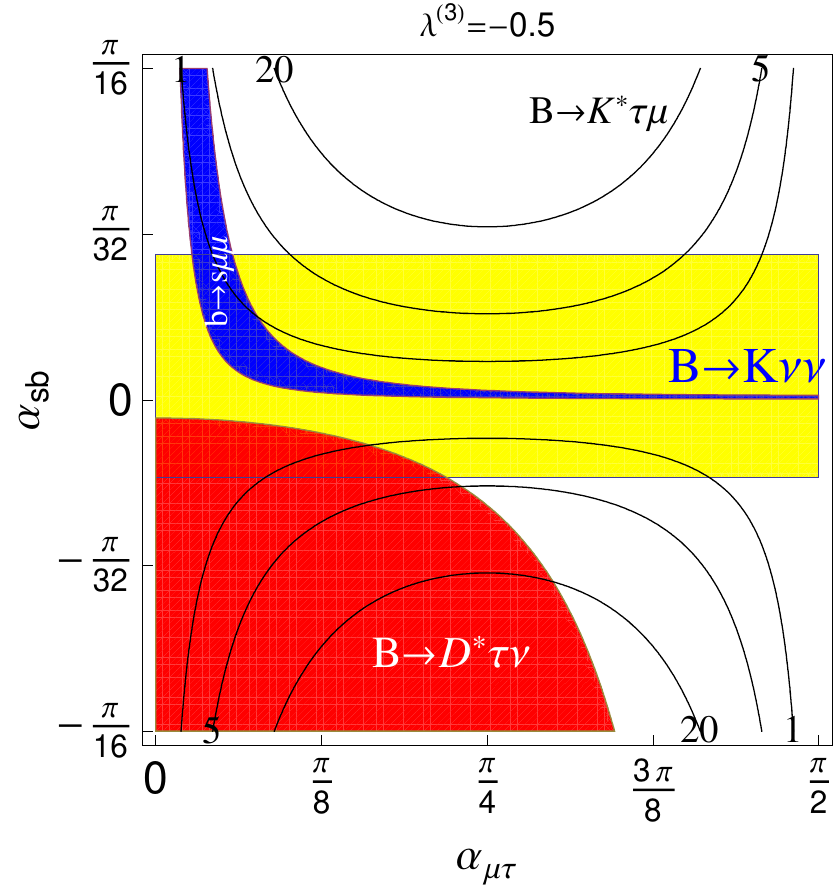}
\includegraphics[width=0.32\textwidth]{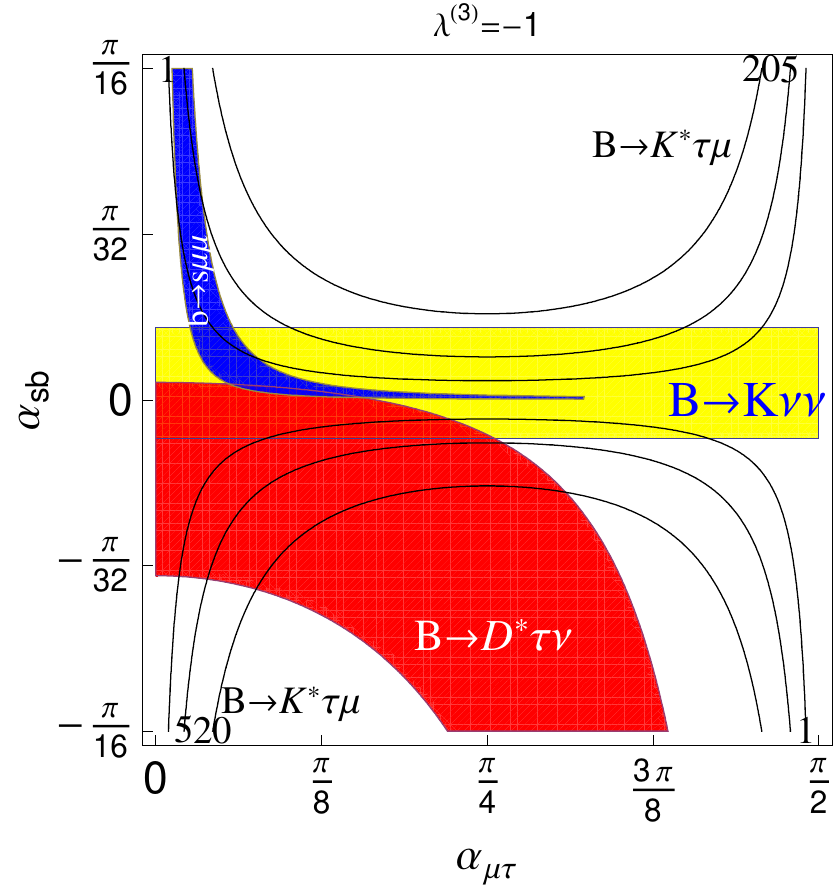}
\includegraphics[width=0.32\textwidth]{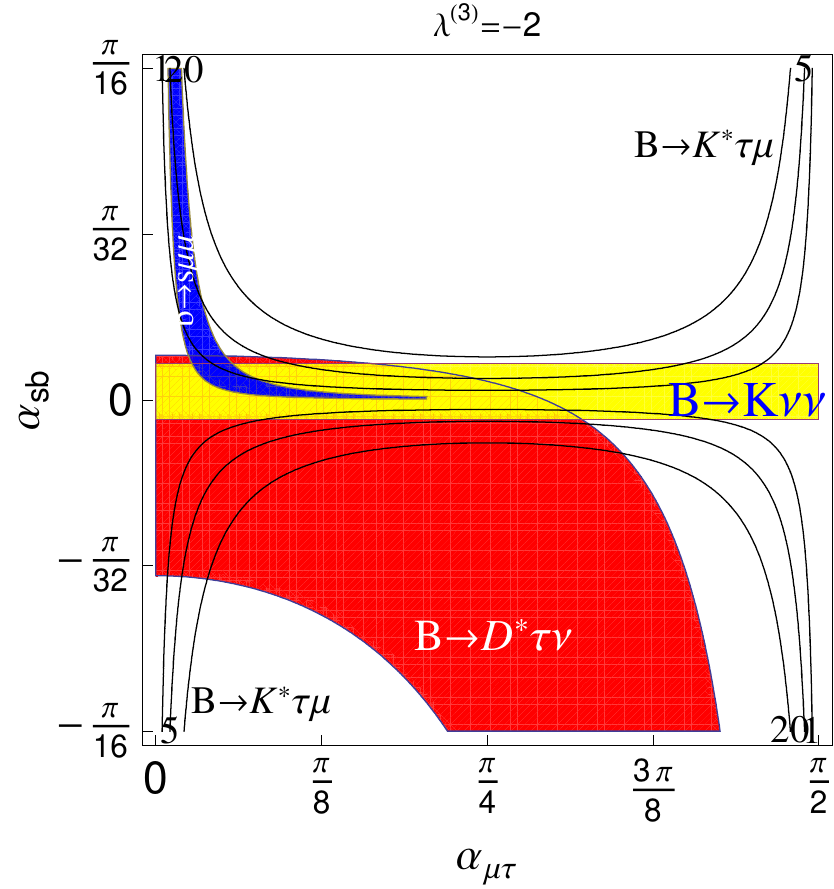}
\end{tabular}
\end{center}
\caption{Allowed regions in the $\alpha_{\mu\tau}$--$\alpha_{sb}$ plane from $B\to K\nu\bar{\nu}$ (yellow), $R(D^*)$ (red) and $b\to s \mu^+\mu^-$ (blue) for $\Lambda=1\,$TeV and $\lambda^{(3)}=-0.5$ (left plot), $\lambda^{(3)}=-1$ (middle) and $\lambda^{(3)}=-2$ (right). Note that $\alpha_{sb}=\pi/64$ roughly corresponds to the angle needed to generate $V_{cb}$ and that if $\lambda^{(3)}$ is positive, $R(D^*)$ and $b\to s\mu^+\mu^-$ cannot be explained simultaneously. \label{Plot2}}
\end{figure*}

We will focus in the following on scenarios with third generation couplings in the EW basis only, which correspond to a general rank 1 matrix in the mass eigenbasis, as suggested in Ref.~\cite{Glashow:2014iga,Bhattacharya:2014wla}. In other words we have
\begin{align}
C_{ijkl}^{(1,3)}&=\lambda^{(1,3)}\tilde X_{ij}\tilde Y_{kl}\,,
\end{align}
\begin{align}
\tilde X &= {L^\dag }XL,\;\tilde Y = {D^\dag }YD\,,\;\;\;
X = Y = \left( {\begin{array}{*{20}{c}}0&0&0\\0&0&0\\0&0&1\end{array}} \right)\,.\nonumber
\end{align}
Taking into account only rotations among the second and third generation one finds
\begin{align}
\tilde X =& \left( {\begin{array}{*{20}{c}}
0&0&0\\
0&{{{\sin }^2}\left( \alpha_{\mu\tau}  \right)}&{ - \sin \left( \alpha_{\mu\tau}  \right)\cos \left( \alpha_{\mu\tau}  \right)}\\
0&{ - \sin \left( \alpha_{\mu\tau}  \right)\cos \left( \alpha_{\mu\tau}  \right)}&{{{\cos }^2}\left( \alpha_{\mu\tau}  \right)}
\end{array}} \right)\,,
\end{align}
\begin{align}
\tilde Y =& \left( {\begin{array}{*{20}{c}}
0&0&0\\
0&{{{\sin }^2}\left( \alpha_{sb}  \right)}&{ - \sin \left( \alpha_{sb}  \right)\cos \left( \alpha_{sb}  \right)}\\
0&{ - \sin \left( \alpha_{sb}  \right)\cos \left( \alpha_{sb}  \right)}&{{{\cos }^2}\left( \alpha_{sb}  \right)}
\end{array}} \right)\,.\nonumber
\end{align}
Note that a rotation $\sin(\alpha_{sb})\gg V_{cb}$ would require fine-tuning with the up sector in order to obtain the correct CKM matrix.

\subsubsection{$Q_{\ell q}^{\left( 1 \right)}$ operator}
In this case we have neutral currents only. As a consequence, there is obviously no effect in $R(D^{(*)})$, but $b\to s \mu^+\mu^-$ is directly correlated to $B\to K^{(*)}\nu\bar{\nu}$ depending on the angle $\alpha_{\mu\tau}$. Note that a change in $\alpha_{sb}$ can be compensated by a change in $\lambda^{(1)}$ and therefore does not affect the correlations among $B\to K^{(*)}\nu\bar{\nu}$ and $b\to s \mu^+\mu^-$ transitions. In Fig.~\ref{Plot1} the regions favoured by $b\to s\mu^+\mu^-$ (blue) and allowed by $B\to K\nu\bar{\nu}$ (yellow) are shown together with contour lines for $B\to K^*\tau\mu$ in units of $10^{-6}$. Note that $B\to K\nu\bar{\nu}$ rules out branching ratios for $B\to K^*\tau\mu$ above approximately $1\times10^{-6}$ and that the constraint from $B\to K\nu\bar{\nu}$, being inclusive in the neutrino flavours, is independent of $\alpha_{\mu\tau}$.

\subsubsection{$Q_{\ell q}^{\left( 3 \right)}$ operator}
Here we have also charged currents that are related to the neutral current processes via CKM rotations. In Fig.~\ref{Plot2} the regions allowed by $B\to K\nu\bar{\nu}$ (yellow) and giving a good fit to data for $b\to s\mu^+\mu^-$ (blue) and (at the $2\,\sigma$ level) for $B\to D^*\tau\nu$ (red) are shown for different values of $\lambda^{(3)}$. Note that $b\to s\mu^+\mu^-$ data can be explained simultaneously with $R(D^{(*)})$ for negative $\mathcal{O}(1)$ values of $\lambda^{(3)}$ without violating the bounds from $B\to K\nu\bar{\nu}$. Again, in the regions compatible with all experimental constraints, the branching ratios of LFV $B$ decays to $\tau\mu$ final states can only be up to $\approx10^{-6}$.

\subsubsection{$Q_{\ell q}^{\left( 1 \right)}$ and $Q_{\ell q}^{\left(3 \right)}$ with $\lambda^{(1)}=\lambda^{(3)}$}
In this case the phenomenology is then rather similar to the case of $C^{(3)}$ only. The major differences are that, as already mentioned before, the bounds from $B\to K\nu\bar{\nu}$ are evaded and the relative contribution to $b\to s\mu\mu$ compared to $R(D^{(*)})$ is a factor of 2 larger. Again $R(D^{(*)})$ rules out very large branching ratios for lepton flavour violating $B$ decays in the regions compatible with $b\to s\mu^+\mu^-$ data. Note that the MFV-like ansatz~\cite{Alonso:2015sja} with additional flavour rotations phenomenologically only slightly differs from the ansatz with third generation couplings.

\subsection{UV completions}
Let us briefly discuss UV completions which can give the desired coupling structure\footnote{While we focus on leptoquarks here, also massive vectors in the triplet representation of $SU(2)$ are a possible UV completion\cite{Greljo:2015mma}.}. As discussed previously, the 4-Fermi operator $Q_{\ell q}^{(3)}$ is relevant both for $R(K)$ and $R(D^{(*)})$. If $Q_{\ell q}^{(3)}$ is mediated by a single field, then there are only four possibilities:
(i) Vector boson (VB) with the SM charges $(SU(3)_{c}, SU(2)_{L}, U(1)_{Y})=({\bf 1},{\bf 3},0)$,
(ii) Scalar leptoquark (SLQ) with ({\bf 3},{\bf 3},$-1/3$),
(iii) Vector leptoquark (VLQ) with ({\bf 3},{\bf 1},2/3),
and (iv) Vector leptoquark with ({\bf 3},{\bf 3},2/3). 
The vector boson ({\bf 1},{\bf 3},0) induces only $Q_{\ell q}^{(3)}$.
On the other hand, the leptoquark fields result in 
particular combinations of $Q_{\ell q}^{(1)}$ and 
$Q_{\ell q}^{(3)}$~\cite{Alonso:2015sja}.
With the assumption of the third generation coupling,
the relative size of the effective couplings $\lambda^{(1,3)}$
and the signs are determined as 
\begin{align}
\text{VB({\bf 1},{\bf 3},$0$) : }& 
\lambda^{(3)}\text{ both positive and negative} ,
\end{align}
\begin{align}
\text{SLQ({\bf 3},{\bf 3},$-1/3$) : }& 
\lambda^{(1)} = 3 \lambda^{(3)},  
 \quad 
\lambda^{(3)}>0,
\\
\text{VLQ({\bf 3},{\bf 1},2/3) : }&
\lambda^{(1)} = \lambda^{(3)},
\quad 
\lambda^{(3)} < 0,
\\
\text{VLQ({\bf 3},{\bf 3},2/3) : }&
\lambda^{(1)} = -3 \lambda^{(3)},
\quad
\lambda^{(3)}
>0.
\end{align}
The coefficient $C_{9}^{ij}$ is proportional to $\lambda^{(1)}+\lambda^{(3)}$ and a negative value is favoured by $R(K)$. Therefore, the scalar leptoquark is rejected as a candidate. To explain $R(D^{(*)})$ simultaneously, $\lambda^{(3)}$ itself must also be negative. This condition excludes the triplet vector leptoquark. If the experimental results are explained by the operator $Q_{\ell q}^{(3)}$ under the assumption of third generation coupling only, the possible mediators are the triplet vector boson or the singlet vector leptoquark.
According to the analysis of the previous section, a good fit to flavour data requires a mediator mass of $\mathcal{O}$(1) TeV. This opens interesting prospects for the LHC, especially in the case of leptoquarks that can be produced in proton-proton collisions via colour interactions and would decay to one lepton ($\tau$ or more interestingly $\mu$) and one jet (possibly a $b$-jet).

\section{Lepton flavour violating $B$ decays}

As lepton flavour universality is violated in $R(K)$ and $B\to D^{(*)}\tau\nu$, and $h\to\tau\mu$ even violates lepton flavour, it is interesting to examine the possibility of observing lepton flavour violating $B$ decays \cite{Glashow:2014iga}. Already from the EFT analysis of the last section it is clear that once gauge invariance is emplyed, LFV $B$ decys cannot be very large if one aims at addressing $R(K)$ and $R(D^{(*)})$ simultaneously. In $Z^\prime$ models the additional constraints from $\tau\to3\mu$ etc. arise. Furthermore, in the UV complete model of Refs.~\cite{Crivellin:2015mga,Crivellin:2015lwa} the branching ratios for LFV $B$ decays are tiny, in general these processes are proportional to $\Gamma_{\mu\tau}\Gamma_{sb}$ and can be large in the presence of sizable flavour violation in the quark and in the lepton sector. Here we review $B\to K^{(*)}\tau^\pm\mu^\mp$ and $B_s\to \tau^\pm\mu^\mp$ in $Z^\prime$ models with generic couplings to fermions \cite{Crivellin:2015era}. $\Gamma^L_{sb}$ can only be large if there are cancellations originating from $\Gamma^R_{sb}$ having the same sign but being much smaller. Therefore, the branching ratios for LFV $B$ decays are bounded by fine tuning together with $\tau\to3\mu$ and $\tau\to\mu\nu\nu$ limiting $\Gamma_{\mu\tau}$. As a result, we find in a scenario in which NP contributions to $C_9^{\ell\ell^\prime}$ only are generated
\begin{align}
{\rm Br}\left[B\to K^{(*)}\tau^\pm\mu^\mp\right]&\le 2.2(4.4)\times 10^{-8}(1+X_{B_s})\,,\\
{\rm Br}\left[B_s\to\tau^\pm\mu^\mp\right]&\le 2.1\times 10^{-8}(1+X_{B_s})\,,
\end{align}
where $X_{B_s}$ measures the degree of fine tuning in the $B_s$ system. Note that these limits are obtained for $\Gamma_{\mu\mu}=0$ (which corresponds to $C_9^{\mu\mu}=0$) and are even stronger for non-vanishing values of $\Gamma_{\mu\mu}$. For $\mu e$ final states the possible branching ratios are much smaller due to the stringent constraints from $\mu\to e\gamma$ and $\mu\to e\nu\nu$.

\section{Conclusion}

In these proceedings we reviewed the impact of the indirect hint for physics beyond the SM in the flavour sector obtained by BABAR, LHCb, CMS and ATLAS on models of NP. We focused on models with $Z^\prime$ bosons and/or additional Higgs doublets but also briefly discussed leptoquarks. While a prime candidate for the explanation of the anomalous $b\to s\mu^+\mu^-$ data is a $Z^\prime$ boson, $h\to\tau\mu$ as well as $B\to D^{(*)}\tau\nu$ can be most naturally explained by an extended scalar sector. Interestingly, models with gauged $L_\mu-L_\tau$ can explain $b\to s\mu^+\mu^-$ data and $h\to\tau\mu$ simultaneously, predicting sizable branching ratios of $\tau\to3\mu$, potentially observable in future experiments. A simultaneous explanation of $b\to s\mu^+\mu^-$ data and $b\to c\tau\mu$ data is possible in the EFT approach with third generation couplings for which a vector leptoquark could be a UV completion\footnote{Rejecting the assumption of third-generation couplings, also a scalar leptoquark would be possible.}.

While the UV complete models \cite{Crivellin:2015mga,Crivellin:2015lwa} predict tiny branching ratios for LFV $B$ decays, these decays can have sizable branching fractions for $\tau\mu$ final states in models with thrid generation couplings and in generic $Z^\prime$ models in the presence of significant fine-tuning in the $B_s-\overline{B}_s$ system.  

\section*{Acknowledgments}
I thank the organizers, for the invitation and the opportunity to present these results. I am grateful to Christian Gross and Stefan Pokorski for useful comments on the manuscript. This work is supported by a Marie Curie Intra-European Fellowship of the European Community's 7th Framework Programme under contract number 
(PIEF-GA-2012-326948).

\bibliography{AMM-tauonic}

\begin{thebibliography}{73}

\bibitem{Egede:2008uy}
U.~Egede, T.~Hurth, J.~Matias, M.~Ramon, W.~Reece, JHEP \textbf{0811}, 032
  (2008), \texttt{0807.2589}

\bibitem{Descotes-Genon:2013vna}
S.~Descotes-Genon, T.~Hurth, J.~Matias, J.~Virto, JHEP \textbf{1305}, 137
  (2013), \texttt{1303.5794}

\bibitem{Aaij:2013qta}
R.~Aaij et~al. (LHCb collaboration), Phys.Rev.Lett. \textbf{111}, 191801
  (2013), \texttt{1308.1707}

\bibitem{Descotes-Genon:2014uoa}
S.~Descotes-Genon, L.~Hofer, J.~Matias, J.~Virto, JHEP \textbf{1412}, 125
  (2014), \texttt{1407.8526}

\bibitem{Altmannshofer:2014rta}
W.~Altmannshofer, D.M. Straub (2014), \texttt{1411.3161}

\bibitem{Jager:2014rwa}
S.~J{\"a}ger, J.~Martin~Camalich (2014), \texttt{1412.3183}

\bibitem{LHCb:2015dla}
T.L. Collaboration (LHCb) (2015)

\bibitem{Aaij:2015esa}
R.~Aaij et~al. (LHCb), JHEP \textbf{09}, 179 (2015), \texttt{1506.08777}

\bibitem{Horgan:2013pva}
R.R. Horgan, Z.~Liu, S.~Meinel, M.~Wingate, Phys.Rev.Lett. \textbf{112}, 212003
  (2014), \texttt{1310.3887}

\bibitem{Horgan:2015vla}
R.~Horgan, Z.~Liu, S.~Meinel, M.~Wingate (2015), \texttt{1501.00367}

\bibitem{Aaij:2014ora}
R.~Aaij et~al. (LHCb collaboration), Phys.Rev.Lett. \textbf{113}, 151601
  (2014), \texttt{1406.6482}

\bibitem{Bobeth:2007dw}
C.~Bobeth, G.~Hiller, G.~Piranishvili, JHEP \textbf{0712}, 040 (2007),
  \texttt{0709.4174}

\bibitem{Altmannshofer:2015sma}
W.~Altmannshofer, D.M. Straub (2015), \texttt{1503.06199}

\bibitem{Descotes-Genon:2015uva}
S.~Descotes-Genon, L.~Hofer, J.~Matias, J.~Virto (2015), \texttt{1510.04239}

\bibitem{Lees:2012xj}
J.~Lees et~al. (BaBar), Phys.Rev.Lett. \textbf{109}, 101802 (2012),
  \texttt{1205.5442}

\bibitem{Huschle:2015rga}
M.~Huschle et~al. (Belle), Phys. Rev. \textbf{D92}, 072014 (2015),
  \texttt{1507.03233}

\bibitem{Aaij:2015yra}
R.~Aaij et~al. (LHCb), Phys. Rev. Lett. \textbf{115}, 111803 (2015), [Addendum:
  Phys. Rev. Lett.115,no.15,159901(2015)], \texttt{1506.08614}

\bibitem{Amhis:2014hma}
Y.~Amhis et~al. (Heavy Flavor Averaging Group (HFAG)) (2014),
  \texttt{1412.7515}

\bibitem{Fajfer:2012vx}
S.~Fajfer, J.F. Kamenik, I.~Nisandzic, Phys.Rev. \textbf{D85}, 094025 (2012),
  \texttt{1203.2654}

\bibitem{CMS:2014hha}
CMS (CMS Collaboration) (2014), {CMS-PAS-HIG-14-005}

\bibitem{Aad:2015gha}
G.~Aad et~al. (ATLAS) (2015), \texttt{1508.03372}

\bibitem{Misiak:2015xwa}
M.~Misiak et~al., Phys. Rev. Lett. \textbf{114}, 221801 (2015),
  \texttt{1503.01789}

\bibitem{Crivellin:2015hha}
A.~Crivellin, J.~Heeck, P.~Stoffer (2015), \texttt{1507.07567}

\bibitem{Krawczyk:1987zj}
P.~Krawczyk, S.~Pokorski, Phys. Rev. Lett. \textbf{60}, 182 (1988)

\bibitem{Crivellin:2013wna}
A.~Crivellin, A.~Kokulu, C.~Greub, Phys.Rev. \textbf{D87}, 094031 (2013),
  \texttt{1303.5877}

\bibitem{Crivellin:2012ye}
A.~Crivellin, C.~Greub, A.~Kokulu, Phys.Rev. \textbf{D86}, 054014 (2012),
  \texttt{1206.2634}

\bibitem{CMS:2013hja}
C.~Collaboration (CMS) (2013)

\bibitem{Branco:2011iw}
G.~Branco, P.~Ferreira, L.~Lavoura, M.~Rebelo, M.~Sher et~al., Phys.Rept.
  \textbf{516}, 1 (2012), \texttt{1106.0034}

\bibitem{Broggio:2014mna}
A.~Broggio, E.J. Chun, M.~Passera, K.M. Patel, S.K. Vempati, JHEP \textbf{11},
  058 (2014), \texttt{1409.3199}

\bibitem{Wang:2014sda}
L.~Wang, X.F. Han, JHEP \textbf{05}, 039 (2015), \texttt{1412.4874}

\bibitem{Abe:2015oca}
T.~Abe, R.~Sato, K.~Yagyu, JHEP \textbf{07}, 064 (2015), \texttt{1504.07059}

\bibitem{Fajfer:2012jt}
S.~Fajfer, J.F. Kamenik, I.~Nisandzic, J.~Zupan, Phys.Rev.Lett. \textbf{109},
  161801 (2012), \texttt{1206.1872}

\bibitem{Sakaki:2013bfa}
Y.~Sakaki, M.~Tanaka, A.~Tayduganov, R.~Watanabe, Phys.Rev. \textbf{D88},
  094012 (2013), \texttt{1309.0301}

\bibitem{Alonso:2015sja}
R.~Alonso, B.~Grinstein, J.M. Camalich, JHEP \textbf{10}, 184 (2015),
  \texttt{1505.05164}

\bibitem{Freytsis:2015qca}
M.~Freytsis, Z.~Ligeti, J.T. Ruderman, Phys. Rev. \textbf{D92}, 054018 (2015),
  \texttt{1506.08896}

\bibitem{Calibbi:2015kma}
L.~Calibbi, A.~Crivellin, T.~Ota, Phys. Rev. Lett. \textbf{115}, 181801 (2015),
  \texttt{1506.02661}

\bibitem{Bauer:2015knc}
M.~Bauer, M.~Neubert (2015), \texttt{1511.01900}

\bibitem{Deshpande:2012rr}
N.G. Deshpande, A.~Menon, JHEP \textbf{01}, 025 (2013), \texttt{1208.4134}

\bibitem{Gauld:2013qba}
R.~Gauld, F.~Goertz, U.~Haisch, Phys.Rev. \textbf{D89}, 015005 (2014),
  \texttt{1308.1959}

\bibitem{Altmannshofer:2014cfa}
W.~Altmannshofer, S.~Gori, M.~Pospelov, I.~Yavin, Phys.Rev. \textbf{D89},
  095033 (2014), \texttt{1403.1269}

\bibitem{Alonso:2014csa}
R.~Alonso, B.~Grinstein, J.~Martin~Camalich, Phys.Rev.Lett. \textbf{113},
  241802 (2014), \texttt{1407.7044}

\bibitem{Hiller:2014yaa}
G.~Hiller, M.~Schmaltz, Phys.Rev. \textbf{D90}, 054014 (2014),
  \texttt{1408.1627}

\bibitem{Ghosh:2014awa}
D.~Ghosh, M.~Nardecchia, S.~Renner, JHEP \textbf{1412}, 131 (2014),
  \texttt{1408.4097}

\bibitem{Crivellin:2015mga}
A.~Crivellin, G.~D'Ambrosio, J.~Heeck (2015), \texttt{1501.00993}

\bibitem{Crivellin:2015lwa}
A.~Crivellin, G.~D'Ambrosio, J.~Heeck (2015), \texttt{1503.03477}

\bibitem{Niehoff:2015bfa}
C.~Niehoff, P.~Stangl, D.M. Straub (2015), \texttt{1503.03865}

\bibitem{Sierra:2015fma}
D.A. Sierra, F.~Staub, A.~Vicente (2015), \texttt{1503.06077}

\bibitem{Celis:2015ara}
A.~Celis, J.~Fuentes-Martin, M.~Jung, H.~Serodio (2015), \texttt{1505.03079}

\bibitem{Belanger:2015nma}
G.~Belanger, C.~Delaunay, S.~Westhoff, Phys. Rev. \textbf{D92}, 055021 (2015),
  \texttt{1507.06660}

\bibitem{Falkowski:2015zwa}
A.~Falkowski, M.~Nardecchia, R.~Ziegler (2015), \texttt{1509.01249}

\bibitem{Carmona:2015ena}
A.~Carmona, F.~Goertz (2015), \texttt{1510.07658}

\bibitem{Buras:2012jb}
A.J. Buras, F.~De~Fazio, J.~Girrbach, JHEP \textbf{1302}, 116 (2013),
  \texttt{1211.1896}

\bibitem{Buras:2013dea}
A.J. Buras, F.~De~Fazio, J.~Girrbach, JHEP \textbf{1402}, 112 (2014),
  \texttt{1311.6729}

\bibitem{Buras:2013qja}
A.J. Buras, J.~Girrbach, JHEP \textbf{1312}, 009 (2013), \texttt{1309.2466}

\bibitem{Gripaios:2014tna}
B.~Gripaios, M.~Nardecchia, S.~Renner, JHEP \textbf{1505}, 006 (2015),
  \texttt{1412.1791}

\bibitem{Becirevic:2015asa}
D.~Becirevic, S.~Fajfer, N.~Kosnik (2015), \texttt{1503.09024}

\bibitem{Varzielas:2015iva}
I.d.M. Varzielas, G.~Hiller (2015), \texttt{1503.01084}

\bibitem{Davidson:2012ds}
S.~Davidson, P.~Verdier, Phys.Rev. \textbf{D86}, 111701 (2012),
  \texttt{1211.1248}

\bibitem{Kopp:2014rva}
J.~Kopp, M.~Nardecchia, JHEP \textbf{1410}, 156 (2014), \texttt{1406.5303}

\bibitem{Falkowski:2013jya}
A.~Falkowski, D.M. Straub, A.~Vicente, JHEP \textbf{1405}, 092 (2014),
  \texttt{1312.5329}

\bibitem{Dorsner:2015mja}
I.~Doršner, S.~Fajfer, A.~Greljo, J.F. Kamenik, N.~Košnik, I.~Nišandžic,
  JHEP \textbf{06}, 108 (2015), \texttt{1502.07784}

\bibitem{Altmannshofer:2015esa}
W.~Altmannshofer, S.~Gori, A.L. Kagan, L.~Silvestrini, J.~Zupan (2015),
  \texttt{1507.07927}

\bibitem{Sierra:2014nqa}
D.~Aristizabal~Sierra, A.~Vicente, Phys. Rev. \textbf{D90}, 115004 (2014),
  \texttt{1409.7690}

\bibitem{Heeck:2014qea}
J.~Heeck, M.~Holthausen, W.~Rodejohann, Y.~Shimizu (2014), \texttt{1412.3671}

\bibitem{Terazawa:1976xx}
H.~Terazawa, K.~Akama, Y.~Chikashige, Phys. Rev. \textbf{D15}, 480 (1977)

\bibitem{Langacker:2008yv}
P.~Langacker, Rev.Mod.Phys. \textbf{81}, 1199 (2009), \texttt{0801.1345}

\bibitem{Aad:2014cka}
G.~Aad et~al. (ATLAS), Phys. Rev. \textbf{D90}, 052005 (2014),
  \texttt{1405.4123}

\bibitem{Buchmuller:1985jz}
W.~Buchmuller, D.~Wyler, Nucl.Phys. \textbf{B268}, 621 (1986)

\bibitem{Grzadkowski:2010es}
B.~Grzadkowski, M.~Iskrzynski, M.~Misiak, J.~Rosiek, JHEP \textbf{1010}, 085
  (2010), \texttt{1008.4884}

\bibitem{Glashow:2014iga}
S.L. Glashow, D.~Guadagnoli, K.~Lane (2014), \texttt{1411.0565}

\bibitem{Bhattacharya:2014wla}
B.~Bhattacharya, A.~Datta, D.~London, S.~Shivashankara (2014),
  \texttt{1412.7164}

\bibitem{Greljo:2015mma}
A.~Greljo, G.~Isidori, D.~Marzocca, JHEP \textbf{07}, 142 (2015),
  \texttt{1506.01705}

\bibitem{Crivellin:2015era}
A.~Crivellin, L.~Hofer, J.~Matias, U.~Nierste, S.~Pokorski et~al. (2015),
  \texttt{1504.07928}

\end{thebibliography}

\end{document}